\documentclass[pre,superscriptaddress,showpacs,twocolumn]{revtex4}

\usepackage{graphicx}
\usepackage{amsmath,amsthm,amssymb}
\usepackage{color}

\begin{document}

\title{Reversible Transport of Interacting Brownian Ratchets}

\date{\today}

\author{Rog{\'e}rio M. da Silva}
\affiliation{Departamento de F\'{\i}sica, Universidade Federal de Pernambuco, 50670-901 Recife, Pernambuco, Brazil}

\author{Cl{\'e}cio C. de Souza Silva}
\affiliation{Departamento de F\'{\i}sica, Universidade Federal de Pernambuco, 50670-901 Recife, Pernambuco, Brazil}

\author{S{\'e}rgio Coutinho}
\affiliation{Departamento de F\'{\i}sica, Universidade Federal de Pernambuco, 50670-901 Recife, Pernambuco, Brazil}

\begin{abstract}
The transport of interacting Brownian particles in a periodic asymmetric (ratchet) substrate is studied numerically. In a zero-temperature regime, the system behaves as a reversible step motor, undergoing multiple sign reversals of the particle current as any of the following parameters are varied: the pinning potential parameters, the particle occupation number, and the excitation amplitude. The reversals are induced by successive changes in the symmetry of the effective ratchet potential produced by the substrate and the fraction of particles which are effectively pinned. At high temperatures and low frequencies, thermal noise assists delocalization of the pinned particles, rendering the system to recover net motion along the gentler direction of the substrate potential. The joint effect of high temperature and high frequency, on the other hand, induces an additional current inversion, this time favoring motion along the direction where the ratchet potential is steeper. The dependence of these properties on the ratchet parameters and particle density is analyzed in detail.
\end{abstract}


\maketitle 

\section{Introduction}

Increasing interest in the manipulation of small objects has attracted much attention to the so called Brownian ratchets~\cite{Magnasco1993, Doering1994, Bartussek1994, Astumian1994,Reimman2002}. These systems are able to promote the directed transport of particles on a periodic substrate lacking reflection symmetry by means of symmetrical, unbiased, non-equilibrium fluctuations. Their applicability can range from understanding the molecular transport of proteins \cite{Julicher1997, Bader1999} to the controlled transport of flux quanta in superconductors \cite{Lee1999,VilegasScience2003,VandeVondel2005,deSouzaSilvaPRB2006,deSouzaSilva2006} and separation of mixtures in nanometric devices \cite{Savelev2003}.
Albeit equilibrium (thermal) fluctuations alone cannot induce such a motion, it plays a key role within transport efficiency when the system is driven out of equilibrium. 

Although these systems are designed with a nominal ``easy'' direction of motion (a gentler slope than that of the ratchet potential), in many cases it is not a trivial task to predict even the real drift direction. In a seminal paper, Doering~\cite{Doering1994} illustrated that when the system is driven by a fast non-equilibrium (but stochastic) fluctuating force, the particle current may be reversed towards the ``hard'' direction, depending upon the noise statistics. Much later, a similar effect was observed for ratchets driven by a high frequency harmonic force~\cite{Bartussek1994}. In this later case, the reversed motion resulted from the interplay between fast driving and equilibrium white noise, as demonstrated analytically by Fistul~\cite{Fistul2002}. In both situations described above, the current reversal was induced by \emph{stochastic} processes.

When several particles are put in a ratchet substrate, their mutual interaction can give rise to a variety of cooperative phenomena~\cite{Derenyi1995,Julicher1995}. Remarkably, even when the substrate potential is perfectly symmetric, directed macroscopic transport of the interacting particles is possible. This can be accomplished, for instance, by spontaneous symmetry breaking when the system is driven out of equilibrium~\cite{Julicher1995} or by inducing correlated fluctuations in the interparticle bonds~\cite{Porto2000}. Interparticle interactions can also induce \emph{deterministic} current reversals. 
A dramatic example is the recently discovered multiple drift reversals of interacting ratchets~\cite{deSouzaSilva2006}. Such an effect was observed experimentally in a system of vortices captured by a nanoengineered periodic array of asymmetric pinning sites in a superconducting film. The vortex drift direction was seen to change sign several times as a function of the vortex occupation number. Numerical simulations suggested that such a sequence of reversals is quite general and results from the interplay between particle-particle and particle-substrate interactions. Indeed, other systems of interacting ratchets seem to exhibit similar behavior~\cite{Pastoriza2005,Reichhardt2007,Gillijns2007}. This reversed current phenomenon has brought about a high level of controllability over the motion of small particles. Although the necessary ingredients seem to be clearly set, the dynamical mechanism behind this effect is not completely understood. Furthermore, the work carried out so far has not explored the influence of thermal noise and non-adiabatic oscillating forces.

In this paper, we set out to illustrate how the drift of long-range interacting small particles on a ratchet substrate can be manipulated by suitably controlling the interplay between particle-particle and particle-substrate interactions, thermal noise, excitation amplitude, and frequency. As demonstrated in Ref.~\cite{deSouzaSilva2006}, the drift of repelling particles undergoes multiple sign reversals, from positive (along the easy ratchet direction) to negative (along the hard ratchet direction) as the number of particles per ratchet period ($n$) changes from odd to even for low amplitude of the external oscillating force and no thermal noise. In the present work, we show that, at a fixed particle density ($n\geq 2$), a sequence of $n-1$ reversals emerges as the excitation amplitude is increased above the ratchet threshold. By detailed characterization of the dynamical states, we conclude that such reversals are induced by dynamical transitions of the particle chain. Furthermore, we undertake a detailed analysis of how thermal fluctuations assist the restoration of the positive current and how the combination of thermal noise and a high excitation frequency leads ultimately to the negative ratchet drift.

\section{System and simulations}

We consider $N$ interacting particles confined in a one-dimensional asymmetric periodic substrate potential and subjected to a fluctuating driving force. We assumed that the particles evolve in time according to the Langevin equation of motion:
\begin{equation}
m\dot{{v}}_{i}  = - \partial_{x_i}\Big({U}_{s}+\sum_{j}{U}_{ij}\Big) - \gamma {v}_{i} + F_d(t) + \xi_i(t),    
\label{Eq.Motion}
\end{equation}	
where $\gamma$ represents the \emph{viscosity} coefficient, $m$ is the particle mass, $U_s$ is the substrate potential, $U_{ij}$ is the pair potential between particles $i$ and $j$, $F_d(t)$ is the fluctuating driving force and $\xi_i(t)$ is the uncorrelated Gaussian noise, which accounts for the thermal fluctuations at a substrate temperature $T$ via the fluctuation-dissipation theorem: $\langle\xi_i(t)\xi_j(t')\rangle = 2\gamma k_BT\delta_{ij}\delta(t-t')$.

\begin{figure}[b,t]
\centerline{\includegraphics[width=0.75\columnwidth]{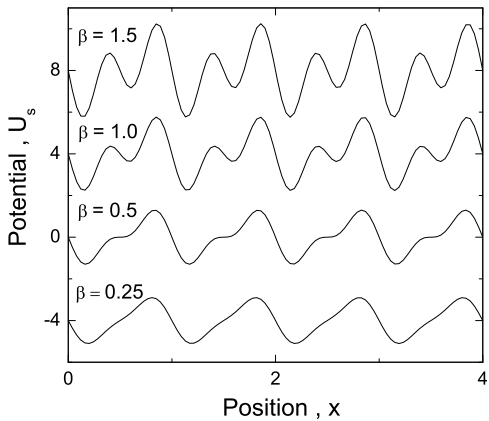}}
\caption{Double-sine potential used in the simulations (Eq.~\ref{Eq.Usub}), for $\beta=0.25$, 0.5, 1.0 and 1.5. The curves are vertically shifted for clarity. Energy in units of the interparticle interaction energy scale, $E_0$, and position in units of the ratchet period, $a$.}
\label{POTENTIALS}
\end{figure}

For the substrate potential, $U_s$, we assumed the usual asymmetric double-sine function~\cite{Reimman2002},
\begin{equation}
{U}_{s}(x_i) = -{U}_{s0} \Big[\sin\Big(\frac{2\pi x_i}{a}\Big)+ \beta \sin\Big(\frac{4\pi x_i}{a}\Big)\Big]. \label{Eq.Usub}
\end{equation}
Here ${U}_{s0}$ and $\beta$ are free parameters controlling the intensity, the asymmetry and the internal structure of the ratchet potential. Some plots of ${U}_{s0}(x)$ are shown in Fig.~\ref{POTENTIALS}.
The shape of the potentials can range from one with a single local minimum, for $\beta\leq 0.5$, to one with two local minima per period, for $\beta>0.5$. The interaction potential between particles $i$ and $j$ is given by a repulsive logarithmic potential, 
\begin{equation}
	U_{ij} = -E_0\ln(|x_i-x_j|).
	\label{Eq.Uvv}
\end{equation}
Such an interaction is relevant for example, for vortices in superconducting thin films. The external driving force is given by $F_d=A\sin(\omega t)$, where $A$ and $\omega$ are the amplitude and the frequency respectively. Hereafter, we shall adopt the following unit system: $E_0$, for energy, $a$, for length, $t_0=\sqrt{ma^2/E_0}$, for time, and $\omega_0=2\pi/t_0$, for angular frequency.

The simulations were performed on a section of the substrate with size $L=10$ and a total number of particles $N$ ranging from 10 to 40 ($n=1,2,3$ and 4) under periodic boundary conditions. The simulations were carried out using a standard finite difference method adapted for stochastic equations. For a given set of parameters, we discarded the first $10^5$ to $10^6$ time steps to allow the system to relax to the dynamical steady state and follow up the same number of time steps for the time averages. To enhance the signal-to-noise ratio, averages were calculated of the physical quantities for several realizations of the thermal noise (up to 50 samples, for low excitation frequencies, and up to 700 samples for high frequencies). The mean particle current, $\langle v\rangle$, is calculated by taking the average of the particle center-of-mass (c.m.) velocities, $v_{\rm cm}=\frac{1}{N}\sum_iv_i$, over time and over the thermal noise realizations. To characterize the dynamical states, the time-averaged density function $\rho(x) = \langle\sum_i\delta(x_i)\rangle$ and the cumulative mass $m(x)=\int_0^x\rho(x')dx'$ were calculated.

To set a few parameters, we chose $m=1$ and $\gamma=16$ in most simulations. Other values of $m$ and a comparison between underdamped and overdamped dynamics is discussed in Sec.~\ref{sec.ov_un}.

\section{Results} \label{Results}

\begin{figure}[b,t]
\centerline{\includegraphics[width=0.98\columnwidth]{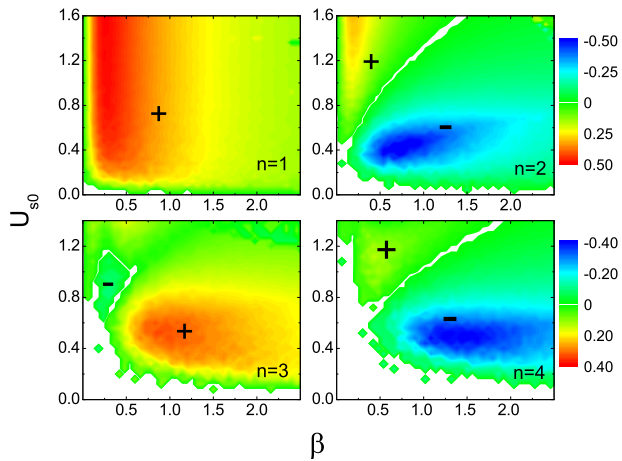}}
\caption{(Color online) Density plots of the (dimensionless) effective asymmetry $\alpha = 1- {{F}_{r}}/{{F}_{l}}$ as a function of the ratchet parameters ${U}_{s0}$ (in units of $E_0$) and $\beta$ (dimensionless), for $n=1$ to $4$. ${F}_{r}$ (${F}_{l}$) is the critical adiabatic depinning force applied to right (left) in the sample. For $\beta \gtrsim 0.5$ and ${U}_{s0} / {E}_{0} \lesssim 1.0$, the current direction is positive for odd $n$ and negative for even $n$.
}
\label{DIAG_COL}
\end{figure}
The deterministic (zero temperature) phase diagrams shown in Fig. \ref{DIAG_COL} indicate the preferred direction of the mean particle current for a given occupation number $n$, ranging from  $1$ to $4$, as a function of $U_{s0}$ and $\beta$, in the limit where the applied force triggers directed motion, i.e. in the ratchet threshold. This measurement was performed by adiabatically increasing an external force until a collective motion of the particles was observed. The measured threshold forces for motion to the right, ${F}_{r}$, and to the left, ${F}_{l}$, were then used to compute the effective asymmetry parameter $\alpha = 1- {F}_{r}/{F}_{l}$, which is positive (negative) for rightward (leftward) motion. The diagrams exhibit the usual sign reversals as a function of $n$ and of the ratchet potential parameters $U_{s0}$ and $\beta$~\cite{deSouzaSilva2006}.

\subsection{The zero-temperature adiabatic regime}
\label{sec.T0}

\begin{figure}[t,b]
\centerline{\includegraphics[width=0.9\columnwidth]{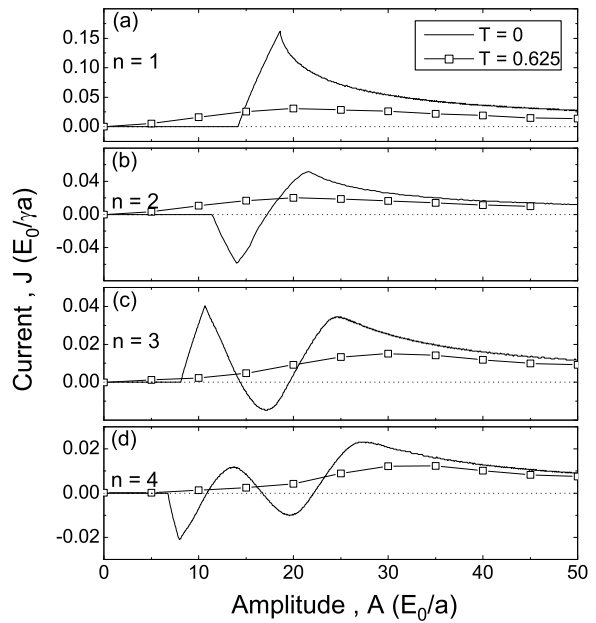}}
\caption{Net particle current $J$ as a function of amplitude $A$ of the sinusoidal drive for the ratchet potential parameters $U_{s0}=0.74$ and $\beta=1.5$. Full lines indicate $T=0$ and line-symbols indicate $T=0.625$, in units of $E_0/k_B$. For $n=1$ (a) there is no current inversion whereas for $n=2$ (b), $n=3$ (c) and $n=4$ (b) respectively, there are clearly $n-1$ current inversions induced by amplitude for the zero-temperature case.}
\label{JxA}
\end{figure}

Here, the ratchet dynamics is explored when the driving force amplitude $A$ is increased within the zero-temperature, low-frequency (adiabatic) regime. Fig.~\ref{JxA}
shows the particle current as a function of $A$ for $n=1$, 2, 3, and 4. The calculations were performed for the ratchet potential parameters $U_{s0}=0.74$ and $\beta=1.5$ (upper curve in Fig.~\ref{POTENTIALS}) and a frequency of $\omega=0.01$. The deterministic ($T=0$) solutions are given by the solid lines. The noisy adiabatic regime will be discussed further on. For $n>1$, $(n-1)$ sign reversals occur with $A$ increasing. Such reversals take place well above the ratchet threshold, contrary to those observed in Fig.~\ref{DIAG_COL}, and therefore are induced by dynamical transitions in the array of particles. 

In order to conduct a detailed analysis of the possible dynamical states and dynamical transitions and how they lead to the observed sign reversals, the dynamics of particles driven by constant force values was simulated. The force intensity was slowly ramped up with the force applied to the positive and then to the negative direction. In this case, the net current is given by $J(F)=J_+(F)+J_-(F)$, with $J_{+(-)}$ being the mean particle current for the force applied to the right (left). Such dc measurements of the ratchet effect is equivalent to an ac measurement where the oscillating force is an adiabatic square wave~\cite{VandeVondel2005}. The advantage of such a strategy is that the dynamical states can be characterized for a constant force during an arbitrary long time interval, thus allowing one to compute statistical quantities for that specific force value. 

\begin{figure}[b,t]
\centerline{\includegraphics[width=0.95\columnwidth]{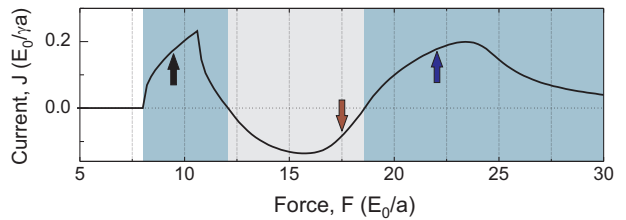}}
\caption{(Color online) Net particle current $J$ as a function of force intensity calculated for a dc measurement of the ratchet effect (see text). Arrows indicate the force intensities used in Fig.~\ref{rho}.} \label{Fig.JxFdc}
\end{figure}
The net current $J$, calculated for the dc measurement described above, as a function of force intensity $F$ for $U_{s0}=0.74$, $\beta=1.5$, and $n=3$, is presented in Fig.~\ref{Fig.JxFdc}. The $J(F)$ curve qualitatively resembles the $J(A)$ curve of Fig.~\ref{JxA}-(c), calculated for sinusoidal drives.

\begin{figure}[b]
\centerline{\includegraphics[width=0.95\columnwidth]{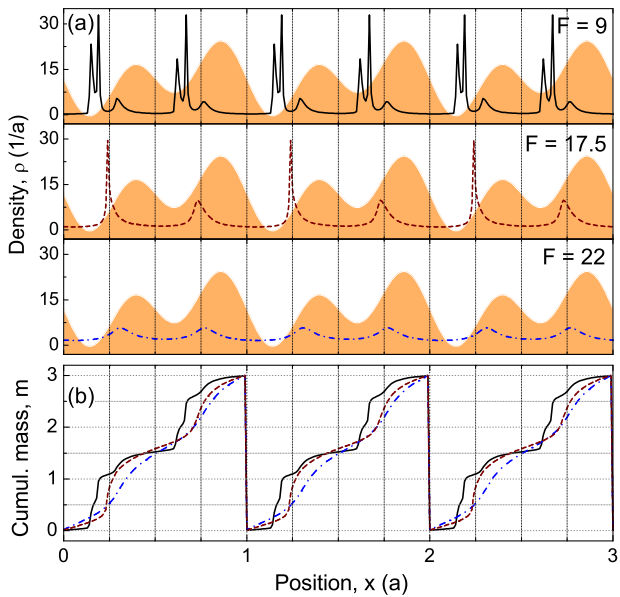}}
\caption{(Color online) Time-averaged particle density $\rho(x)$ (a) and (dimensionless) cumulative mass $m(x)$ (b) for positive, constant drive values $F=9$ (full lines), 17.5 (dashes), and 22 (dash-dots). The background area plots represent the substrate potential used in this calculation.}
\label{rho}
\end{figure}

Fig.~\ref{rho} presents plots of the particle density $\rho(x)$ (panel (a)) and of the cumulative mass $m(x)$ (panel (b)) for $n=3$ and positive driving forces with intensities $F=9$, 17.5, and 22. As indicated in Fig.~\ref{Fig.JxFdc}, these force intensities correspond to 
low-amplitude positive current, negative current, and high-amplitude positive current, respectively. 
The region of low-amplitude positive drift is characterized by a double peak at each local minimum of the substrate potential, which is represented in Fig.~\ref{rho} as the background area plots. The jumps in $m(x)$, integrated for $F=9$, is evidence that each double peak observed in $\rho(x)$ corresponds to exactly one particle distributed in the close vicinity of the substrate local minima, i.e. two  \emph{localized} particles per substrate period, while the smoother regions of $m(x)$ correspond to a third particle spread out across each substrate period, i.e. one \emph{delocalized} particle per substrate period. This leads to the conclusion that all minima of the substrate potential are occupied most of the time, except for a very short time interval during which a particle trapped in one minimum is knocked out and replace by another. Therefore, the effective dynamics may be roughly described as $N/3$ (\emph{delocalized}) particles moving over an effective ratchet potential composed of the bare substrate potential $U_s$ plus the potential due to the other $2N/3$ (\emph{localized}) particles fixed at the local minima of the substrate potential. Clearly, this effective potential preserves the symmetry of the substrate potential, favoring the observed positive current sign. For $F=17.5$, however, the $\rho(x)$ peaks at the weaker minima are drastically reduced and the mass distributes broadly near these minima in comparison to the sharp distribution near strong minima. Now, the effective potential comprises $N/3$ particles localized at the stronger minima, rendering an inversion of the effective asymmetry of the system. Finally, at $F=22$, all $\rho(x)$  peaks are small and there is no point in the substrate potential where $m(x)$ is sharply distributed. Hence all particles are now delocalized and the system restores its original symmetry. 

\begin{figure}[t,b]
\centerline{\includegraphics[width=0.95\columnwidth]{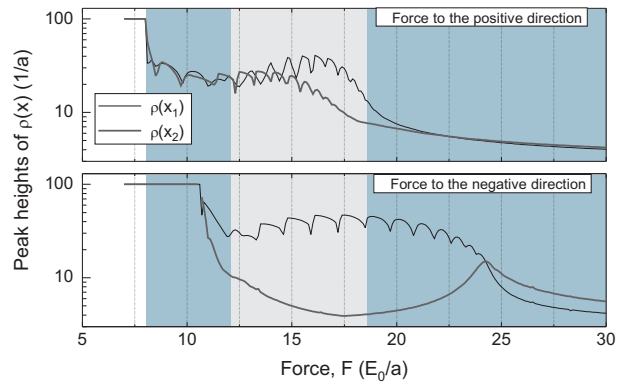}}
\caption{(Color online) Local maximum values of the time-averaged particle density as a function of force intensity for positive (top) and negative (bottom) drive directions. The maxima were measured in the vicinity of the strong (thick lines) and the weak (thin line) potential wells (see Fig.~\ref{rho}).}
\label{rhopeaks}
\end{figure}

For the negative drive direction, a similar sequence of dynamical transitions was observed, but with different critical values of $F$. An overall picture of what happens at positive and negative drive directions and the relevance for the observed ratchet states is given in Fig.~\ref{rhopeaks} 
where the main peaks located at the deeper ($\rho(x_1)$) and the shallower ($\rho(x_2)$) local minima are plotted against force intensity for different drive orientations. Roughly speaking, the negative current region is characterized most of the time by only one localized particle per substrate period, which tends to invert the system symmetry inducing the negative current, while the positive current regions are mostly characterized by either two (low amplitude) or no (high amplitude) localized particles, which tends to keep the system symmetry. The exact force intensities where the sign inversions take place are a non-trivial result from the competition between the dynamics at positive and negative drive direction.

More generally, at small drive amplitudes, for any $n>1$, there are $(n-1)$ particles in a substrate period that can be considered to be localized whereas one particle is delocalized. As the amplitude increases, the localized particles are dynamically delocalized one by one, inducing a total of $(n-1)$ sign reversals. In other words, the system is sequentially reduced to dynamical states similar to those observed at the small amplitude regime of lower particle density systems.

\subsection{Noise-induced sign inversions}

To check the effect of thermal fluctuations in the adiabatic limit, detailed simulations of Eq.~\ref{Eq.Motion} have been performed for $T\geq 0$ and low frequency ($\omega=0.01$). Fig.~\ref{JxA} presents the mean particle current calculated for $T=0.625$ (line-dot), which corresponds to the strong noise limit. For this temperature value, the ratchet current is essentially positive for all $n$, particularly in the amplitude ranges where the current direction at $T=0$ is negative, that is, for such amplitude ranges, \emph{thermal noise induces a current reversal}. 

\begin{figure*}[t]
\centerline{\includegraphics[width=1.80\columnwidth]{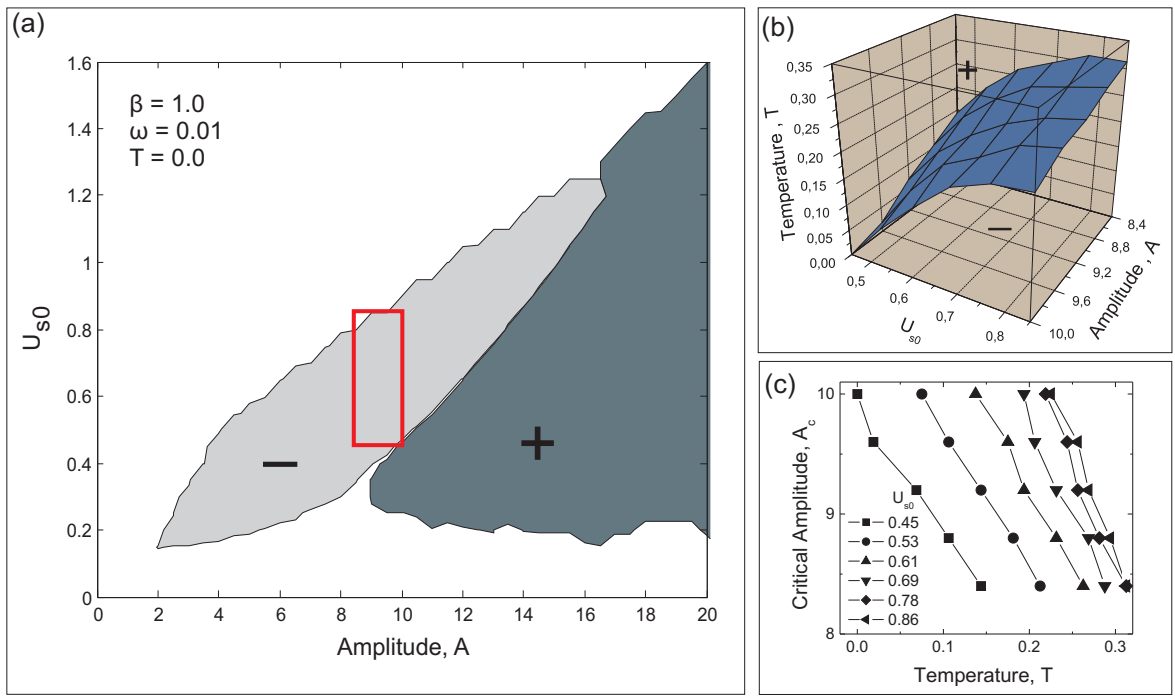}}
\caption{(Color online)(a) Zero-temperature phase diagram of the drift direction of a $n=2$ chain for $\beta=1.0$, $\omega=0.01$. (b) 3D phase diagram of the same system showing the inversion temperature surface separating the negative (low temperature) and positive (high temperature) current regions as a function of ${U}_{s0}$ and $A$. The ranges of ${U}_{s0}$ and $A$ are represented by the rectangle in (a). (c) Temperature dependence of the critical amplitude, $A_c$, where thermally assisted inversions take place, calculated for several $U_{s0}$.}
\label{ThermalRev}
\end{figure*}

To investigate these noise-induced inversions in greater detail, we calculated the temperature dependence of the net current, $J(T)$, of the $n=2$ particle array for $\omega=0.01$, $\beta=1.0$, and several values of the substrate strength $U_{s0}$ and the drive amplitude $A$. Fig.~\ref{ThermalRev} 
(a) maps the region in the $U_{s0}$-$A$ plane where the $T=0$ drift state is negative. The rectangle depicts the region considered for the $T>0$ simulations. The results of the $J(T)$ sweeps were used to build the $T$-$U_{s0}$-$A$ diagram shown in Fig.~\ref{ThermalRev} (b), where the surface represents the transition temperature, $T_{inv}(A,U_{s0})$, between the negative and the positive drift states.

Interestingly, $T_{inv}(A)$ seems to decrease linearly with the amplitude. In other words, the critical amplitude, $A_c$, at which the thermally-induced drift inversion occurs, scales as $A_c=a-bT$.  This becomes evident in Fig.~\ref{ThermalRev}-(c) where $A_c(T)$ is plotted for several values of $U_s0$. To understand this behavior, we first recall the results of the previous paragraphs, which show that the inverted ratchet drift observed for a $n=2$ particle array can be described as half of the particles moving over the potential generated by the substrate and the remaining (localized) particles. Therefore, one can roughly associate $A_c$ with the force where such dynamics is destroyed, that is, where the localized particles delocalize, thus restoring the original substrate symmetry. On the other hand, the linear behavior of $A_c(T)$ suggests a transition to a regime of creep motion resulting from thermal activation of the localized particles over the energy barriers of the (tilted) substrate potential. In systems exhibiting similar dynamics, such as flux lines in hard superconductors~\cite{Kim62} and domain walls in ferromagnetic materials~\cite{Lemerle98}, the critical force for a detectable creep motion decreases linearly with $T$ as long as the relevant energy barrier at zero drive, $\Delta U_s$, has no explicit temperature dependence and $k_BT\ll\Delta U_s$~\cite{Anderson64}. In the present case, for $\beta=1.0$, $\Delta U_s \sim 4U_{s0}$, which is much bigger than $k_BT$ near the transition surface, as can easily be inferred from Fig.~\ref{ThermalRev}-(b).

\subsection{The high frequency regime}

\begin{figure}[b]
\centerline{\includegraphics[width=0.85\columnwidth]{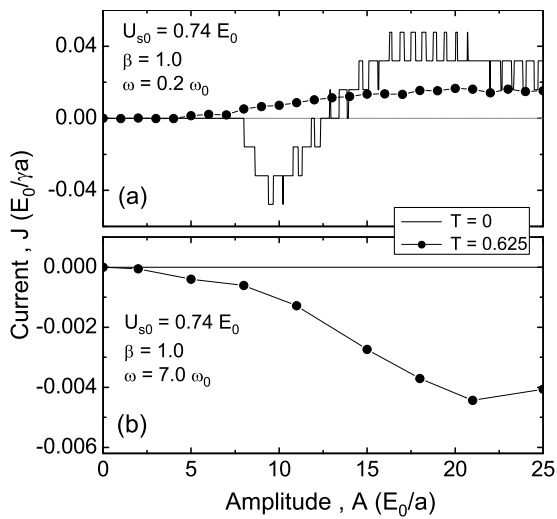}}
\caption{Mean particle current of the $n=2$ chain as a function of amplitude for (a) moderate frequency, $\omega=0.2$, and (b) high frequency, $\omega=7.0$. Substrate parameters are $\beta=1.0$ and ${U}_{s0}=0.74$. Full lines indicate $T=0$ and symbol-line plots indicate $T=0.5$ (a) and $T=0.625$ (b) in units of $E_0/k_B$.}
\label{HighFreq}
\end{figure}

Here we turn our attention to the effects of high frequency drives. To fix some quantities, we set $n=2$, $U_{s0}=0.74$, and $\beta=1.0$. The amplitude dependence of the mean current for a few frequency and temperature values is shown in Fig.~\ref{HighFreq}. 

At $T=0$, $J(A)$ is characterized by a series of plateaus and spikes produced by the well-known lock-in effect observed in single-particle rocking ratchets. Notice, however, that the mean particle velocity assumes discretized values at multiples of $v_0 = 0.5\omega a/2\pi=0.016$, which means that each particle moves across an average distance of $a/2$ during a cycle of the external drive. This result is in contrast with that of single-particle ratchets, where the lock-in state corresponds to the particle moving a step forward across a full period of the substrate in one drive cycle, which gives $v_0=\omega a/2\pi$~\cite{Bartussek1994}. By performing similar simulations for other particle occupation number values, we have observed that, in general, $v_0=\omega a/2\pi n$, corresponding to an average particle displacement of $a/n$ per drive cycle. This result provides the opportunity for controlling the particle current accurately, at a given frequency, by changing the particle density. 

At non-zero low temperatures the lock-in effect is smeared out by thermal fluctuations. At sufficiently high temperatures, thermally induced current reversals, similar to those observed in the adiabatic regime, are observed in the amplitude range where the zero-temperature drift direction is negative (Fig.~\ref{HighFreq}(a)). However, when the frequency is even higher, the combination of fast rocking ($\omega=7$) and thermal fluctuations ($T=0.625$) leads to another reversal, resulting in the negative current sign in the full amplitude range. This result is somewhat expected since, as mentioned above, at a high $T$ the particles may be considered independent of one another and, for non-interacting rocking ratchets at high temperatures, a sufficiently high driving frequency induces the negative current~\cite{Bartussek1994, Fistul2002}.

\begin{figure}[b,t]
\centerline{\includegraphics[width=0.85\columnwidth]{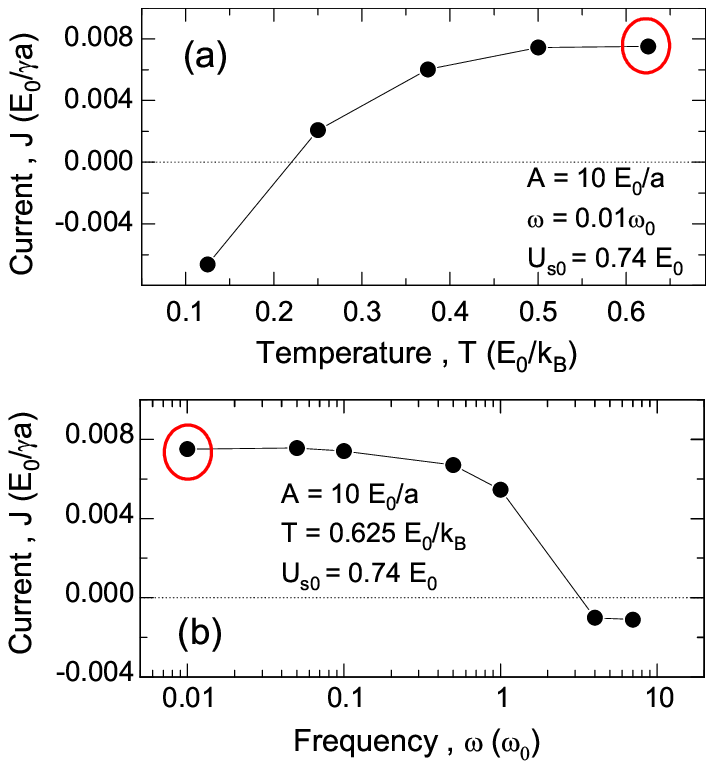}}
\caption{(Color online) (a) Mean particle current, $J$, of the $n=2$ chain as a function of temperature, $T$, in the adiabatic regime ($\omega=0.01$) for $\beta=1.0$, ${U}_{s0}=0.74$, and $A=10$.  (b) Frequency dependence of $J$ for the same substrate parameters and $A=10$ starting from the point indicated by a circle in (a) ($T=0.625$). }
\label{INVERSAO_FREQ}
\end{figure}
Fig. \ref{INVERSAO_FREQ} illustrates the sequence of sign inversions when one first increases the temperature in the adiabatic ($\omega=0.01$) regime (a) and then, at a fixed high temperature value ($T=0.625$), increases the frequency up to $\omega=10$ (b). Note that the current induced by high frequencies is approximately one order of magnitude smaller than that induced by temperature. Indeed, the frequency-induced reversed current of non-interacting ratchets is usually small because such inversions occur at frequencies much higher than the natural frequency of the system. The same behavior is observed here because at a high frequency and temperature the system is essentially non-interacting. Interacting ratchets in the adiabatic regime, on the other hand, offer several ways to reverse the sign of the current (by changing either the particle density, the force amplitude or the temperature) while keeping its absolute value at the same order of magnitude.

\subsection{Overdamped versus underdamped dynamics}
\label{sec.ov_un}

\begin{figure}[b,t]
\centerline{\includegraphics[width=1.00\columnwidth]{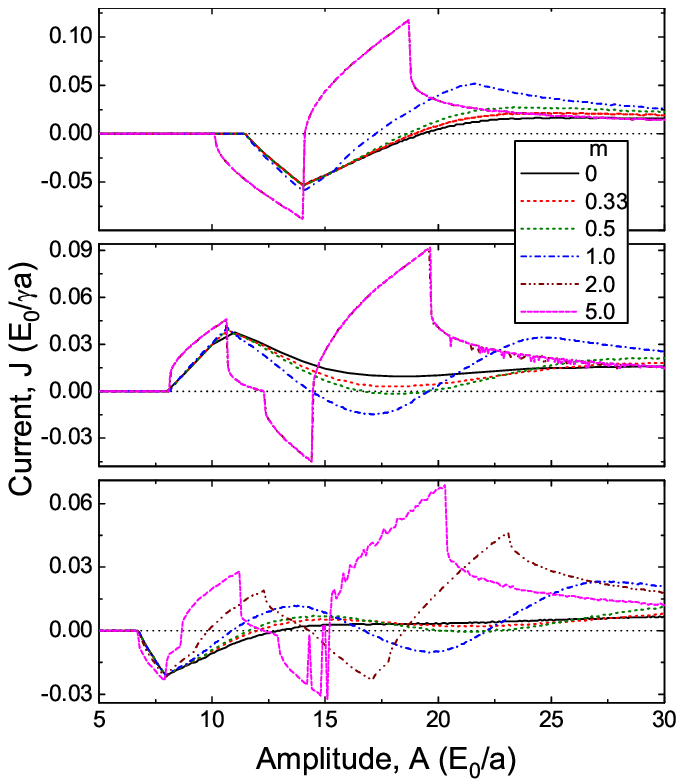}}
\caption{(Color online) Net particle current $J$ as a function of amplitude drive $A$ for $n=2$ (a), $n=3$ (b) and $n=4$ (c) calculated for different mass values: $m=0$ (full lines), 0.33 (dashes), 0.5 (short-dashes), 1.0 (dash-dots), 2.0 (dash-double-dots) and 5.0 (long dashes). The ratchet parameters are the same as in Fig.~\ref{JxA}.}
\label{underXover}
\end{figure}

In the previous paragraphs, we explored the dynamical properties of 1D arrays of particles of unity mass moving on a ratchet substrate with friction coefficient $\gamma = 16$. Here, we shall discuss other values of $m$, spanning the overdamped and underdamped regimes. Although the ratchet threshold diagrams calculated for the different $m$ values are very similar to those presented in Fig.~\ref{DIAG_COL} for $m=1$, the behavior of the moving phases are qualitatively different for different intervals of $m$ values. 

As noticed by Borromeo et al. \cite{Borromeo2002}, the particle current (for rocking, non-interacting ratchets) behaves qualitatively different from the usual overdamped ($m=0$) limit when the ratio $\gamma/m$ becomes smaller than the libration frequency of the particles at the bottom of the ratchet wells ($\omega_{\rm lib}=\sqrt{k/m}$, with $k$ the spring constant of the well). In this moderate underdamped regime, the onset of the ratchet current becomes abrupt and the ratchet tail in $\langle J\rangle(A)$ decays faster as compared to the $\langle J\rangle(A)$ of an overdamped ratchet. When $\gamma/m\ll\omega_{\rm lib}$, however, the above described behavior is substituted by a scenario dominated by chaotic dynamics.

In our case, with the ratchet potential given by Eq.~\ref{Eq.Usub} and choosing $\beta=1.5$, one obtains $k=262U_{s0}/a^2$, so that the critical mass that roughly separates the underdamped and overdamped regimes is $m^*=\gamma^2/k=1.32$. Fig.~\ref{underXover} shows the $\langle J\rangle(A)$ characteristics for particle densities $n=2$, 3 and 4, and different mass values, ranging from 0 to 5. Larger $m$ values lead to chaotic dynamics, which is out of the scope the present paper (in fact, already for $m=5.0$,  chaotic trajectories were observed for a few drive amplitude values resulting on the fluctuations exhibited by the $\langle J\rangle(A)$ curves for  $m=5.0$ and $n=4$). As it is clear, for $m>m^*$ (i.e. $m=2.0$ and $m=5.0$), the transitions between different ratchet phases are more abrupt than those observed for $m\leq 1.0$.  Notwithstanding, the $n-1$ oscillations are still present and the behavior of $\langle J\rangle(A)$ for $m=2.0$ and $m=5.0$ is quite revealing with respect to the scenario described in Sec.~\ref{sec.T0}. For instance, the interval $A>8.7$ in the $n=4,m=5.0$ curve seems to qualitatively imitate the whole $n=3,m=5.0$ curve, whereas the interval $A>12.3$ in the $n=3,m=5.0$ curve seems to qualitatively imitate the ($n=2,m=5.0$) curve. This is consistent with the fact that it is the number of localized particles which determines the ratchet direction. The qualitative similarity between curves with different $n$ arises from the fact that in the underdamped regime the ratio between the particle current and the force amplitude has a universal shape which does not depend on the details of the potential but rather on its critical forces~\cite{Borromeo2002}.

For $m<m^*$, the curves are much smoother and, for $n=3$ and $n=4$, the high-amplitude negative drift phase becomes positive for $m < 0.5$. Notwithstanding, the oscillations, reminiscent of the interplay between localized and delocalized particles, are still present, even in the $m=0$ case.

\section{Summary}

In summary, the transport properties of interacting particles confined in a 1D ratchet potential has been investigated in detail. It has been shown that, in the low-temperature adiabatic regime, the particle chain of density $n$ undergoes a series of $n-1$ current inversions as the oscillating drive amplitude is increased for a range of the investigated parameters. A systematic study of the time-averaged particle density and cumulative mass, calculated at different drive intensities, suggests that these inversions result from the competition between different dynamical states of the chain where a certain fraction of the particles are effectively trapped by the substrate, while the others float over the effective potential generated by the substrate and the localized particles. The number of localized particles determines the symmetry of the effective ratchet potential and, consequently, the preferential drift direction. In the non-zero thermal noise regime, the oscillating behavior of the amplitude-dependent mean current observed at low temperatures gives place to a positive-drift phase in the whole amplitude range at a sufficiently high temperature. In cases where the drift is negative at $T=0$, the current direction becomes positive at a certain temperature-dependent critical amplitude $A_c(T)$, which decreases linearly. Such behavior suggests that these noise-induced inversions can be associated with thermal activation of localized particles over the substrate barriers. Finally, the high-frequency regime has been addressed for both weak and strong noises. For weak noise, the steps and spikes reminiscent of the well-known lock-in effect have been observed. But here the size of the steps can be fine-tuned by changing the particle density. In the strong noise case, the interplay between thermal fluctuations and the fast oscillations of the driving force leads to the complete inversion towards the negative drift direction in the whole amplitude range.

\bibliography{BrownianMotors}

\end{document}